\documentclass[useAMS,usenatbib]{mn2e} 
\usepackage{graphicx}
 
\def\solmas{$\mathrm{M_\odot}$~}
\def\solmasp{$\mathrm{M_\odot}$}

\def\tff{$t_{ff}$}

\def\rhooph{$\rho$ Oph}
\def\be{\begin{equation}}
\def\ee{\end{equation}}

\title[Clump Lifetimes and the IMF]
{Clump Lifetimes and the Initial Mass Function}
\author[Clark, Klessen \& Bonnell]
{Paul C. Clark$^1$ \thanks{E-mail: pcc@ita.uni-heidelberg.de}, 
Ralf S. Klessen$^1$
\& Ian A. Bonnell$^2$ \\
$^1$ Institut f\"ur Theoretische Astrophysik, Universit\"at Heidelberg,
Albert-Ueberle-Str. 2, Heidelberg, Germany  \\
$^2$ School of Physics \&
Astronomy, University of St Andrews, North Haugh, St Andrews, Fife, KY16 9SS, UK 
}

\date{\today}

\begin{document}
\maketitle

%
% Abstract....
%

\begin{abstract}

Recent studies of dense clumps/cores in a number of regions of low-mass star
formation have shown that the mass distribution of these clumps closely resembles
the initial mass function (IMF) of field stars. One possible interpretation of
these observations is that we are witnessing the fragmentation of the clouds into
the IMF, and the observed clumps are bound pre-stellar cores. In this paper, we
highlight a potential difficulty in this interpretation, namely that clumps of
varying mass are likely to have systematically varying lifetimes. This `timescale'
problem can effectively destroy the similarity bewteen the clump and stellar mass
functions, such that a stellar-like clump mass function (CMF) results in a much
steeper stellar IMF. We also discuss some ways in which this problem may be
avoided.

\end{abstract}

%
% Key words
%

\begin{keywords}
stars: formation - stars: pre-main-sequence - stars: mass function - ISM: clouds
\end{keywords}

%
% Paper contents...
%

\section{Introduction}
\label{Introduction}

One of the major goals in studying star formation is to understand what
determines the observed distribution of stellar masses, the initial mass function
(IMF). There have been a number of ideas to explain the origins of the IMF
including fragmentation, accretion, magnetic fields and stellar feedback (for a
list of theories, see references within \citealt*{Shuetal1987};
\citealt{MacLowKlessen2004}; \citealt*{Bonnelletal2005b}). One of the most
intriguing developments has been the recent finding that the mass distribution of
prestellar cores, those that have not yet, but appear to be on the verge of
forming stars, is similar to the stellar IMF (\citealt*{Motteetal1998,
TestiSargent1998};\citealt{Johnstoneetal2000,Johnstoneetal2001,
Johnstoneetal2006}; \citealt*{NutterWardThompson2006}).  This has led to the suggestion that fragmentation
during the pre-stellar regime leads directly to the stellar IMF. In such a cloud
fragmentation picture, the IMF is essentially primordial, since the
clump/core masses are presumed to provide the main gas reservoir for each forming
system. Furthermore, models of turbulent fragmentation may be able to explain
this distribution of clump masses lending additional weight to this scenario
\citep*{Fleck1982, HunterFleck1982, Elmegreen1993, Padoan1995, Padoanetal1997,
Myers2000, Klessen2001, PN2002}.

In this paper, we highlight a problem in interpreting the observed clump mass
distribution as a population of bound, prestellar, cores. We show that such an
interpretation would suggest a final stellar IMF which is significantly steeper
than the observed mass function. In Section \ref{observations} we briefly discuss
the observational studies that have examined the clump mass distributions in the
nearby star forming regions. Section \ref{timescales} contains our description of
the timescale problem that one encounters when assuming that these clumps are to be
the progenitors of individual stars or systems, and we extend this discussion to
include significant fragmentation in Section \ref{multiplemj}. In Section
\ref{avoidance} we discuss ways in which the timescale problem can be avoided and
we summurise the paper in Section \ref{conclusions}. In this paper, we will use the
term `clump' to refer to any density enhancement, and `core' or `prestellar' core
to refer to those clumps which are going to form stars. We also assume in this
paper that the clumps have a density contrast with respect to the ambient cloud
such that they are gravitationally decoupled from their surroundings.

\section{Clump/Core Observations in Low Mass Star Forming Regions}
\label{observations}

The first large study of clump masses was published by \citet{Motteetal1998}, for
a population of submillimetre cores in \rhooph. Using data obtained with IRAM,
they discovered a total of 58 starless clumps, ranging in mass from 0.05~\solmas
to $\sim 3$~\solmasp. The clump mass function (CMF) was shown to be similar to
that of the stellar IMF, with power law fit $dn/dm \propto m^{-\alpha}$ following
$\alpha = 2.5$ above $\sim0.5$~\solmas and  $\alpha = 1.5$ below.
\citet{TestiSargent1998} confirmed that clumps in Serpens also have a similar
mass distribution, taken with the OVRO millimeter-wave array, with their clumps
following $\alpha = 2.1$ between $\sim 0.6$~\solmas and $\sim 20$~\solmas
(although some potential caveats with this method have been discussed by
\citealt{Ossenkopfetal2001}). 

Further studies by Johnstone and colleagues of clump properties have been
conducted with SCUBA on the JCMT, focusing on \rhooph~\citep{Johnstoneetal2000},
the Orion B North region \citep{Johnstoneetal2001} and the Orion B south region
\citep{Johnstoneetal2006}. The 55 clumps from the \rhooph~study were found to
cover a slighly larger range than the \citet{Motteetal1998} study, going from
0.02~\solmas to 6.3~\solmasp. The mass spectrum was however very similar, with
$\alpha = 1.5$ below 0.6~\solmas and $\alpha = 2.5$ above, despite differences in
both the observational methods and the clump finding techniques used. While
observations of clumps in Orion B North yield similar results to those in Serpens
and \rhooph, there does appear to be a distinct difference in the clump
properties in Orion B South. The 57 identified cores in the
\citet{Johnstoneetal2006} study span a mass range from $\sim 0.4$~\solmas to
$\sim 28$~\solmas but have a turnover from $\alpha = 2.5$ to $\alpha \sim 1.5$ at
somewhere between 3 - 10~\solmasp, clearly a much higher turnover mass than that
found in the other regions. \citet{NutterWardThompson2006} have taken this
further, using SCUBA archive data on Orion, and find that this higher turnover is
true for the region in general. A simlar result was found in the Pipe nebula (see
\citealt*{Ladaetal2006}), using extinction mapping. One can find more complete
discussions of the properties of clumps, or `cores', in the reviews of
\citet{DiFrancescoetal2007} and \citet{WardThompsonetal2007}.

An exciting interpretation of these observations is that we are witnessing the
direct formation of the IMF via fragmentation of the parent cloud. This suggests
that there could be a mapping between the observed clumps and the final IMF, with the
clumps providing the primary reservior of material for each stars or sytems that
form within.  Clearly this is an enticing picture, since it implies that one may
be able to directly study the origin of the IMF, simply by examining the
observable features of the gas. Also, the fact that clumps extend smoothly down
past the hydrogen burning limit may then suggest that brown dwarfs form as part
of the same process as general star formation \citep{PadoanNordlund2004}.
\citet*{Greavesetal2003} have even discovered a potential preplanetary clump,
which may suggest that this cloud fragmentation process can extend down into the
planet mass regime.

However, the evidence that these clumps are the direct origin of the stellar IMF
is by no means conclusive.  In particular, Johnstone and collaborators
\citep{Johnstoneetal2000,Johnstoneetal2001,Johnstoneetal2006} demonstate that the
clumps in their studies are more consistent with stable Bonner-Ebert spheres, and
are thus unlikely to be in a state of active star formation. This appears to be
the case for all the regions they have studied, including \rhooph, in which
\citet{Motteetal1998} claim that the majority of clumps are bound. Until
recently, most of studies have little, or no, line-width information and are thus
unable to accurately determine the internal thermal and kinetic energies of the
clumps. Some work has been done for \rhooph~\citep*{Bellocheetal2001,
Andreetal2007} and NGC~2068 in the Orion B cloud (Andr{\'e}, {\it{private
communication}}), which shows that the velocities are low in these regions,
suggesting that the contribution to support from internal rotation, and/or
turbulence, may be small and so the clumps are bound. These detailed molecular
line surveys will we be able to address the concerns that we raise in this paper.

\begin{figure}
\includegraphics[width=3.4in,height=3.4in]
{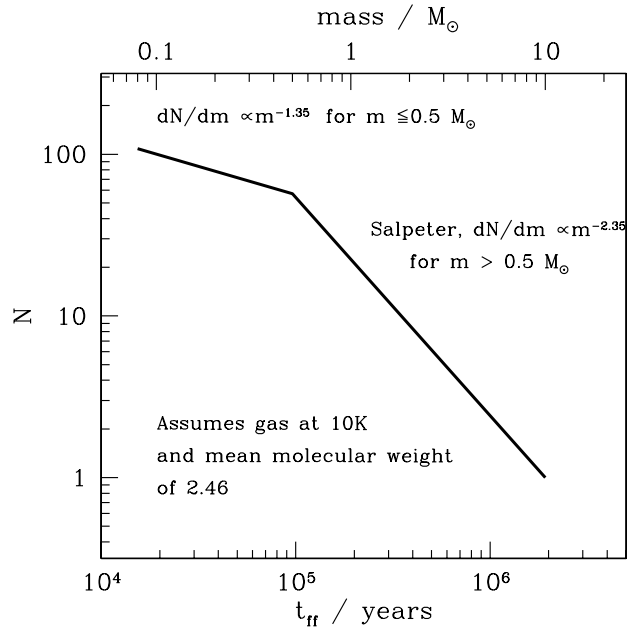}
\caption{\label{timescaleplot}
We plot here the range in free-fall times, \tff, for a
simple IMF, assuming each object in the IMF originates from a clump 
which has one Jeans mass. The free-fall times are calculated from the
Jeans density using a gas temperature of 10~K and a mean molecular weight
of $\mu = 2.46$.}
\end{figure}

\section{The timescale for fragmentation}
\label{timescales}

\begin{figure*}
\centerline{ 	\includegraphics[width=3.4in,height=3.4in]
{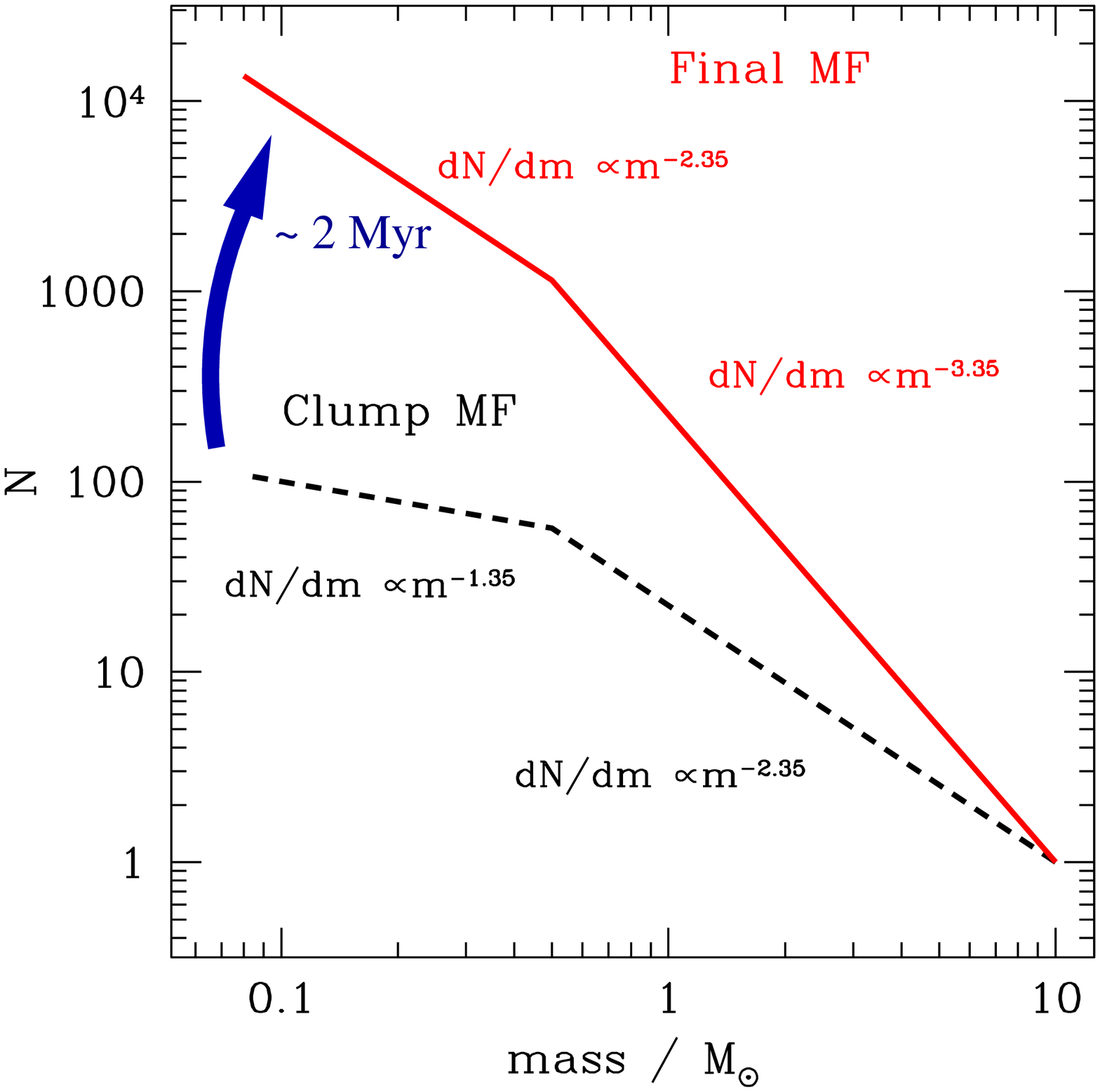}
              \includegraphics[width=3.4in,height=3.4in]
{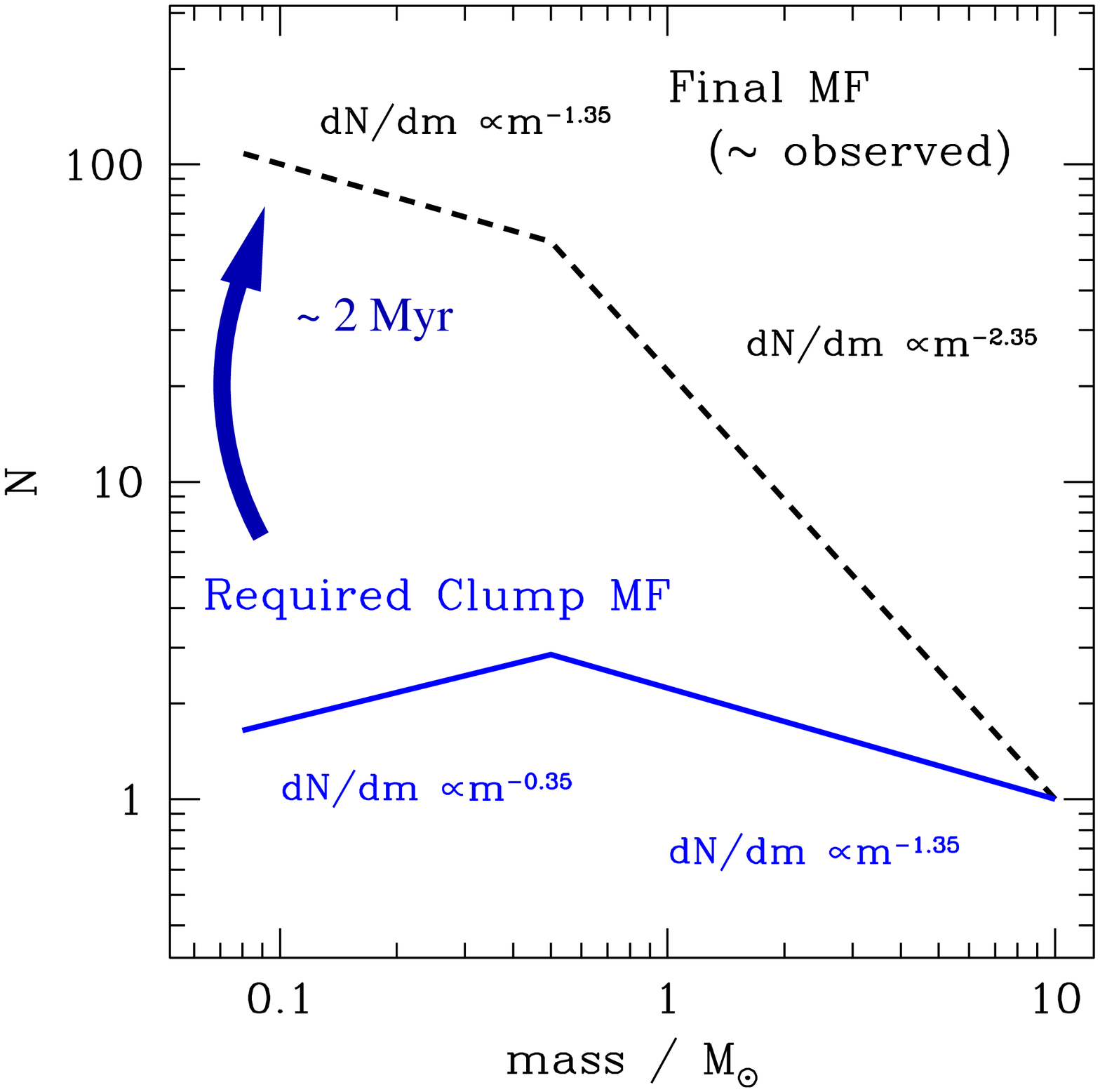}}
\caption{\label{imfevol} In this figure, we demonstrate what happens when a clump
population is  permitted to be constantly replenished and for each clump to
collapse to form 1 star on its free-fall time (so assuming that each clump has
only 1 Jeans mass). The free-fall time for each clump is the same as those given
in Figure \ref{timescaleplot}. and the time over which the clump population is
evolved is just that for the most massive  clump present to collapse, which for a
10 \solmas star is $t_{ff}(10~M_{\odot}) \sim$~2~Myrs.  In the left-hand plot, we
start with a clump mass population that is very similar to the IMF, which is
represented here by the dashed line. The solid line shows the resulting mass
spectrum of protostars, after 2~Myrs, taking our three assumptions above. In the
right hand plot we show, with the solid line, the clump mass spectrum required to
produce a population of protostars which is roughly the same as the stellar IMF.
See Section \ref{timescales} for a discussion of these graphs.} 
\end{figure*}

The purpose of this paper is to demonstrate that mapping the observed clump mass
function onto the stellar IMF is not straightforward. The problem lies in the fact
that clumps of different mass will most likely evolve on different timescales. If
one denotes the evolution time of a clump by $t_{evol}$, then the final IMF,
$f_{IMF}(m)$, is related to the CMF, $f_{CMF}(m)$, by, 

\be
\label{cmfimfmap}
f_{IMF}(m) \approx \frac{f_{CMF}(m)~\tau_{CMF}}{t_{evol}(m)}
\ee

\noindent where $m$ denotes the mass of the objects in each case, and $\tau_{CMF}$
is the timescale over which the CMF exists (and we suggest below that this is
typically longer than $t_{evol}(m)$, for most $m$). If the evolution timescale for
the clump is dependent on its mass, then any power law form that is taken for the
CMF will differ from the form of the IMF.

In this section, we will illustrate this problem using some simple Jeans mass and
timescale arguments. For the timescale, in the following discussion, we will use
the free-fall time, \tff, since we are discussing the idea that these clumps are
the bound progenitors of young, protostellar, systems.

If each clump is to collapse to form a star, or small $N$ system, then it must
have at least one Jeans mass by definition \citep{Jeans1902}. In its simplest
form, the Jeans mass can be thought of as the critical mass at which the
(negative) gravitiational energy exceeds the internal energy. For a uniform
density sphere, this corresponds to a critical mass,

\be
\label{jeansmass}
m_{J} = \left[ \frac{4\pi \rho}{3} \right]^{-1/2} 
\left[ \frac{5}{2}\frac{kT}{G\mu}\right]^{3/2},
\ee

\noindent where $\rho$ is the density, $T$ is the temperature, $\mu$ is the mean
molecular weight, and $k$ and $G$ are the Boltzmann and gravitational constants,
respectively. For simplicity, we will first assume that each clump has one Jeans
mass, although we will relax this later on. Assuming the temperature in the
region is roughly constant, the clump mass then needs to vary with the clump
density in the same way as the Jeans mass,

\be
m \propto \rho^{-1/2},
\ee

\noindent such that low mass clumps need to have a higher density than their high
mass companions. Note that a single density is sufficient to describe the clump,
since gravity only responds to the volume averaged density (for example, the
critical Bonnor-Ebert mass is only a factor of $\sim 2$~smaller than the Jeans
mass, despite the density profile in the Bonnor-Ebert sphere). In the absence of
any other form of support, and assuming no further external compression, the core
will collapse on its free-fall time, \tff, which is again dependent on the volume
averaged density,

\be
t_{ff} \propto \rho^{-1/2}.
\ee

\citet{WardThompsonetal2007} have also shown that there exists a systemic trend in
clump lifetime with density, and at the densities we consider here, this is of the
order of the free-fall time. If the clumps are to be marginally bound, then their
collapse timescale should be a function of their mass,

\be
\label{collapsetime}
t_{collapse} \propto m_{clump}.
\ee

Even in cases where magnetic fields and ambipolar diffusion are significant, once
the clump becomes bound (supercritical), it collapses on a timescale which is
proportional to the free-fall time \citep{TassisMouschovias2004}. So if one
assumes that all the clumps are involved in creating stars, higher mass clumps
should take longer to form their protostars than lower mass clumps. Given the two
orders of magnitude in mass that is present in the IMF, one then requires a
population of clumps which collapse with a similar range in timescales. The clump
mass function can then be converted to a distribution of collapse timescales,
which we plot in Figure \ref{timescaleplot}. In making this figure, we have used
the Jeans mass, given by equation \ref{jeansmass} and the free-fall time, given
by

\be
\label{freefalltime}
t_{ff} = \left( \frac{3 \pi}{32 G \rho} \right)^{1/2},
\ee

\noindent where $G$ is the standard gravitational constant. In calculating the
density from the Jeans mass, we assume a gas temperature of 10K and a mean
molecular weight of 2.46.

There are two immediate questions. First, assuming that these clumps are
the progenitors of future stellar systems, and noting that they should have
different evolutionary timescales, why do we see all the clumps at the
same time? Second, why do we also see this same distribution in each of the
nearby star-forming regions?

The fact that (roughly) the same clump mass distribution is seen in so many
regions of active star formation suggests that such a clump population is
constant in time, since all the regions have different ages. Otherwise we would
have to assume that we have caught these regions at a very special time in their
evolution, and that that special time is right now. This is made more unlikely by
the fact that both the dynamical timescale (or crossing time) and the free-fall
time of the low mass clumps are remarkably short (as little as $10^{4}$ years) in
comparison to the age estimates of local star-forming regions. Therefore `now'
would have to be within $10^{4}$ years of each region's special evolutionary
phase. It thus seems reasonable to assume that the observed form of the clump
mass function is not a brief phase in a star-forming region's evolution. It
should also be noted here that the clump mass function is also expected to be
time-independent in an environment which is dominated by driven turbulence  (for
example, see \citealt{BallesterosParedesetal2006,Padoanetal2007}).

The mapping between the clumps and stars/systems can only make sense if what we
observe as the clump mass distribution is a uniquely occurring population of
pre-stellar cores, and we show here why this is the case. If the clump population
is constantly collapsing to form stars on local free-fall times, and if the
clumps are constantly being replenished, then one would have a mass function of
stars that has many more low mass objects than what is observed for the IMF. This
is because low mass clumps have shorter lifetimes than their higher mass siblings, and
they are being constantly replenished such that the pre-collapse clump population
remains constant in time.

We can demonstrate this by evolving a simple mass function for the time that the
highest mass clump takes to collapse, and making the following assumptions:

\begin{enumerate}
\item The clump mass distribution is constant in time (always replenished).
\item The clumps each have roughly 1 Jeans mass.
\item The clumps collapse on the corresponding free-fall timescale.
\end{enumerate}

For illustration, we will assume a clump IMF of

\be
\label{IMF}
\begin{array}{ll}
dN(m) \propto m^{-1.35}   	dm & 0.08 < m / M_{\odot} < 0.5 	\\
				   & 					\\
dN(m) \propto m^{-2.35}   	dm & m / M_{\odot} > 0.5, 		\\ 
\end{array}
\ee

\noindent which is similar to those quoted in Section \ref{observations} which
have been observed in active regions of nearby star formation. Note here that our
high mass power law is roughly Salpeter \citep{Salpeter1955}, but neither the
exact form of the mass function, nor the lowest mass bin, are important to the
following discussion.

\noindent In Figure \ref{imfevol}, we plot the evolution of two clump mass
funtions. The left hand plot shows how a clump population, that is similar to the
stellar IMF, evolves to form a population of protostars, based on the three
assumptions stated above. Since the highest mass clump in the mass distribution
has 10~\solmasp, we evolve the clump mass spectra until this clump has collapsed,
that is, for a time of roughly 2~Myrs. In Expression \ref{collapsetime}, we see
that there exists a linear relationship between the timescale for collapse and
the mass of the clump. This means that ten 1~\solmas stars can form in the same
period as one 10~\solmas star. Such a linear timescale-mass relationship then
results in an increase in the power of the final mass function, such that a clump
mass distribution described by Expression \ref{IMF} results in a much steeper
stellar IMF like, 

\be
\label{IMFresulting}
\begin{array}{ll}
dN(m) \propto m^{-2.35}   	dm & 0.08 < m / M_{\odot} < 0.5 	\\
				   & 					\\
dN(m) \propto m^{-3.35}   	dm & m / M_{\odot} > 0.5.		\\
\end{array}
\ee

\noindent This is clearly much steeper than the stellar IMF, and not consistent
with the clump IMF that we start with.

Using this result, it is then possible to work backwards, and determine what
clump mass function one should observe if such a process is to generate the IMF.
This is done by simply by subtracting 1 from the power in the mass function,
giving,

\be
\label{IMFrequired}
\begin{array}{ll}
dN(m) \propto m^{-0.35}   	dm & 0.08 < m / M_{\odot} < 0.5 	\\
				   & 					\\
dN(m) \propto m^{-1.35}   	dm & m / M_{\odot} > 0.5. 		\\
\end{array}
\ee

\noindent This is shown graphically in the right hand plot in Figure
\ref{imfevol}. It is clear for the figure that the required clump mass function
is much shallower than those quoted in Section \ref{observations}, for the simple
picture we have outlined here. In fact, the required clump mass function is much
more similar to that commonly quoted for the internal structure of molecular
clouds in general, which is observed to follow a $\sim 1.5$ power
\citep*{Loren1989,StutzkiGuesten1990,Blitz1991,Williamsetal1994}.

\section{Clumps with Multiple Jeans Masses}
\label{multiplemj}

In the previous section we showed how a distribution of clump masses would evolve
into a stellar population, assuming that the clump distribution was constant in
time and that each clump had 1 thermal Jeans mass. In such conditions, each clump
would form roughly 1 star, resulting in a strong mapping between the clump mass
function and the IMF (as has been suggested by \citealt{Shuetal2004}). This is an
extreme picture. Current observations show that the multiplicity of young stellar
objects is higher than in the field-star population, and that these protostellar
systems exist on scales smaller, or similar to, the average clump size in the
region \citep{Ducheneetal2004,Correiaetal2006}. Indeed, there is now mounting
evidence that the observed clump mass distribution should be the origin of small
$N$ systems rather than single stars (for example, see \citealt{Andreetal2000,
Goodwinetal2004a, Goodwinetal2004b}). For the clumps to fragment, it is likely
that they will then need to have serveral Jeans masses in their initial
configuration \citep{Tohline1980, Larson1985, Shuetal1987, Bastienetal1991,
BurkertBodenheimer1993, Bonnell1999, TsuribeInutsuka1999,Tohline2002, 
Sterziketal2003, Goodwinetal2004a, Goodwinetal2004b}. One therefore needs to
examine the evolution of a clump mass distribution in which the members have a
range of Jeans masses. 

\begin{figure}
\includegraphics[width=3.4in,height=3.4in]
{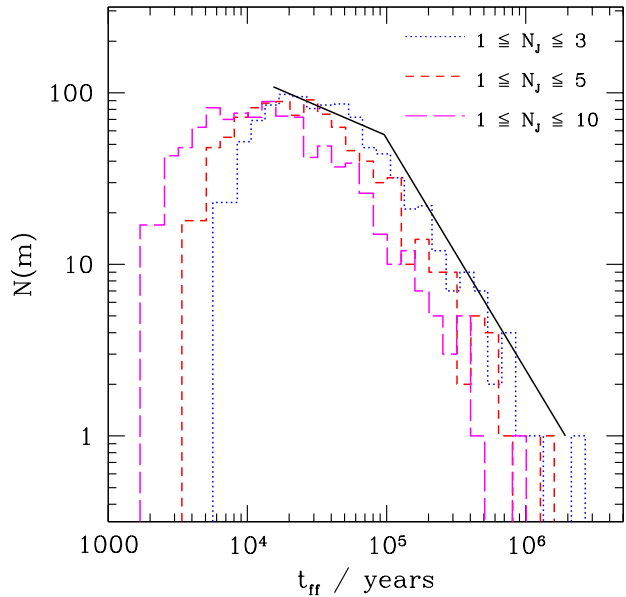}
\caption{\label{tffdist}
We plot here the range in free-fall times, \tff, for a
simple clump mass function (same as in figure \ref{timescaleplot}), but
this time we assume that the clumps have a random number of Jeans masses,
within some predefined range. The assumption of 1 Jeans mass per clump
is represented by the solid mass function. 
The free-fall times are calculated from the
Jeans density using a gas temperature of 10~K and a mean molecular weight
of $\mu = 2.46$.}
\end{figure}

In Figure \ref{tffdist} we plot the timescale distribution for a clump
distribution which covers the same mass range as is given in Expression
\ref{IMF}. Assuming that each clump has one Jeans mass, we get the analytic
solution discussed above. This is shown in Figure \ref{tffdist} as the solid
(black) line. Note that this has the exactly the same form as that plotted in
Figure \ref{timescaleplot}. We also plot 3 other distributions in Figure
\ref{tffdist}, in which we assume each clump can have between 1 and $N_{J}$
Jeans masses, with $N_{J} =$ 3, 5 and 10. For each clump, $N_{J}$ is chosen
randomly from a uniform distribution in the permitted range. A clump with mass
$m_{clump}$ will then have an associated Jeans mass of $m_{J} =
m_{clump}/N_{J}$. If the temperature remains constant, the density of the clump
is then given by the Jeans mass relation (\ref{jeansmass}), which is turn
translates into a free-fall time, given by Equation \ref{freefalltime}. It is
the distribution of these new timescale that is plotted in Figure
\ref{tffdist}.

The important feature to note from Figure \ref{tffdist} is that the
distribution of collapse timescales becomes wider as the clumps increase their
potential to fragment. This means that clumps which are capable of forming
small $N$ systems, such as binaries or triples, are also susceptable to the
timescale problem that we discuss in Section \ref{timescales}. 

\section{Avoiding the timescale problem}
\label{avoidance}

The most obvious way the clump population could avoid the timescale problem is
for all the clumps to have roughly the same density, since then their free-fall
timescales would be centred around a regional value. 

This is the route \citet{PN2002} follow. They describe how an IMF-like
distribution of Jeans unstable clumps can be formed naturally via the
supersonic turbulence in molecular clouds. They predict that the shock
jump-conditions in a molecular gas give rise to a clump population which is
characterised by a roughly uniform density (Note that $\rho_{clump} \propto
m_{clump}^{1/6}$ in their model). The higher-mass clumps are, most likely,
Jeans unstable, since the Jeans mass is lower for these clumps. However as
one moves to progressively lower-mass clumps, this is not the case. One
must then look to what fraction of the total clumps have a density higher
than that predicted from the jump-conditions, and \citet{PN2002} make use
of the density PDF to estimate this bound fraction of clumps. As a result,
the high-mass clumps in the Salpeter section are characterised by single
density, and so the largest have many Jeans masses, while the lower-mass
clumps follow the same density scaling law as we describe above. The
lower-mass clumps (to the left of the turnover point) are therefore
susceptable to the timescale problem that we outline in this paper. We
therefore expect that the \citet{PN2002} CMF would result in more low-mass
stars than is seen in the stellar IMF.

While a single common clump density does get around the timescale problem, it is
unlikely that the CMF will be the dominant factor which controls the IMF under
these conditions. This is due to the process of competitive accretion
(\citealt{Larson1978, Zinnecker1982, Bonnelletal1997}; \citealt{Bonnelletal2001a,
Bonnelletal2001b}; \citealt*{Klessen2001, Bonnelletal2007}; although see also the
debate between \citealt*{Krumholzetal2005} and \citealt{BonnellBate2006}), which
will result in the clumps competing for the available mass, including the mass
which currently `belongs' to other clumps. A single density also means a single
Jeans mass (assuming a roughly constant temperature), and so clumps masses
progressively larger than this Jeans mass become increasingly more unstable to
fragmentation \citep{BurkertBodenheimer1993, TsuribeInutsuka1999, Donateetal2004,
Goodwinetal2004a, Goodwinetal2004b}. This results in small $N$ systems, in which
competitive accretion will also be an important process. 

Thus it seems that while the timescale problem can be avoided if the clumps are
characterised by a single density, competetive accretion would most likely
dominate the final IMF. For a CMF to be the progenitor of the IMF, the majority of
the objects must be bound, have a common density and be able to resist both high
levels of sub-fragmentation and competitive accretion. Note that recent results
show that even weak magnetic fields may be able to inhibit fragmentation to some
extent \citep{HoskingWhitworth2004,Ziegler2005,Fromangetal2006}, so the
fragmentation issue might still be avoided.

One could perhaps argue that the discussion in Section \ref{timescales} above
is too simplistic, since we ignored any non-thermal forms of support, such as
rotation, turbulence and magnetic fields, all of which could potentially alter
the timescale for the collapse of a clump. However from numerical studies it is
typically found that rotational or turbulent support does not substantially
alter the collapse timescale (for example, see \citealt{TsuribeInutsuka1999};
\citealt{Banerjeeetal2006}). The reason is simply that such forms of support
are typically non-isotropic, allowing collapse to occur along favorable
directions, with a timescale given by (roughly) the free-fall time. A similar
argument holds for magnetic support \citep{Heitschetal2001}.

In a magnetically dominated gas it is theoretically possible to remove the mass
dependency of the clump collapse timescale, providing both the local
mass-to-flux ratio and ionisation fractions are favourably balanced
\citep{TassisMouschovias2004}. However, at present, there is no compelling
reason why such a delicate balance should exist. One can construct a similar
situation by varying the temperature in the clumps. In such a case, the Jeans
mass variation from clump to clump would be controlled primarily by the
temperature, with the density being constant ($m_{J} \propto
\rho^{-1/2}T^{3/2}$). However this would still require a temperature range from
around 2~K to 45~K for clumps in the range 0.1~\solmas to 10~\solmasp,
respectively. Line-width observations suggest that such a wide range of
temperatures is not to be expected, and indeed clump mass estimates from sub-mm
observations tend to assume a single temperature for a star-forming `core' in
the range 10 - 20~K.

The `hierarchical' fragmentation picture \citep{Hoyle1953} would also avoid the
timescale problem, and indeed \citet{Larson1973} suggested that such a process may
be able to form a mass function of fragments similar to the IMF.  One could
perhaps argue that the regions in which the clump masses are observed to be larger
\citep{Johnstoneetal2006, Ladaetal2006, NutterWardThompson2006} are just in an
earlier stage of the hierarchical collapse process. However as \citet{Larson1973}
pointed out, one needs some way of preventing the fragments from merging and in
fact no simulation to date has shown signs of hierarchical fragmentation, as it is
described in the models \citep{Larson2007}. The main problem is that it is not
possible for density fluctuations to grow faster than the ambient collapse, unless
these fluctuations have Jeans mass to start with \citep{Tohline1980}, which
results in initial conditions that have many Jeans masses. These are exactly the
conditions required for the competitive accretion process
\citep{BonnellBate2006}. 

The timescale problem is avoided entirely if the clump mass population plays no
part in shaping the IMF, and the similarities between the two mass function is
just a coincidence. Recent numerical simulations have suggested that this might be
the case. In both studies of driven \citep{Klessenetal2000,Klessen2001} and freely
decaying turbulence \citep{ClarkBonnell2005}, the vast majority of clumps are
unbound, with only the more massive of those in the distribution gaining the
neccessary conditions to form stars. Not only do the lower mass clumps have
kinetic energies in excess of their gravitational potential energy, but they very
seldom possess a Jeans mass (that is, their thermal energy is larger that that
from self-gravity, for example see \citealt{Klessenetal2005}). The highest mass
objects, forming at the stagnation points of convergent turbulent flows,
eventually become completely bound, gain multiple Jeans masses and fragment into
small $N$ groups. The IMF in such simulations is largely controlled by the
competitve accretion process. The fact that there exists an ever present
population of clumps which resembles the IMF is never a problem, since almost all
the low mass objects are transient \citep{Klessen2001, TilleyPudritz2004,
ClarkBonnell2005, VazquezSemadenietal2006, ClarkBonnell2006}.

Note that the timescale problem that we highlight here applies to all theories
which directly try to produce an IMF-style distribution of fragments, with
exception of the `hierarchical' model \citep{Hoyle1953, Larson1973}, although
as we note above, this contains its own problems. This is also true (in varying
degrees) of the `turbulent fragmentation' theories \citep{Fleck1982,
HunterFleck1982, Elmegreen1993, Padoan1995, Padoanetal1997, Myers2000, PN2002},
and \citet{Elmegreen1993} raised a similar point to the one we make here in his
study,

\section{Conclusions}
\label{conclusions}

We have presented a potential problem with interpreting the observed clump mass
function (CMF) as the direct origin of the stellar (or system) IMF. If each clump
is assumed to be a star-forming core, then it must have at least one Jeans mass.
If the clumps have a comparable number of Jeans masses, then the low-mass clumps
must have much higher densities than the high-mass clumps. This in turn translates
into a range of free-fall times which are proportional to clump mass. Thus clumps
of different mass are evolving on different timescales. If one then assumes the
clump mass function is constant in time, as is suggested by its presence in most
of the nearby star forming regions, then the resulting stellar mass function is
significantly steeper than the observed IMF (an increase of $+1$~in the power-law
fit). 

The alternative is that all clumps have the same density, and thus evolve on the
same timescale. However this also means that the more massive clumps are perhaps
increasingly susceptible to fragmentation and may produce produce systems of
lower-mass stars (although see the recent magnetic studies which yield reduced
levels of fragmentation:
\citealt{HoskingWhitworth2004,Ziegler2005,Fromangetal2006}). Attaining higher
masses would then necessitate subsequent accretion. Under such conditions we point
out that competitive accretion should become an important ingredient in shaping
the final IMF (for example, see \citealt{BonnellBate2006}).

We therefore conclude that the mass function of clumps that directly turns into
individual stars and higher-order systems, needs to be either considerably
shallower than is inferred for nearby star-forming regions, or needs to avoid
sub-fragmentation and competitive accretion.

\section{Acknowledgements}

The authors would like to thank the LOC responsible for the `Early Phase of Star
Formation' meeting held at Ringberg Castle Germany (28th August to 1st September
2006) which provided the final impedous for the completion of this paper. On a
more personal level, we would also like to thank Phillipe Andr\'e, Bruce
Elmegreen, Richard Larson, Derek Ward-Thompson, Frank Shu, Robi Banerjee, Ant
Whitworth and the anonymous referees for many helpful discussions. PCC
acknowledges support from the German Science Foundation (via grant KL1358/5).

\bibliographystyle{mn2e}
\bibliography{/home/homer/0/pcc/bibfile/pccbib}

\end{document}